\documentclass[usenatbib,referee]{mn2e}
\usepackage{graphicx}

\def\tE{t_{\rm E}}
\def\Dd{D_{\rm d}}
\def\Ds{D_{\rm s}}
\def\dDd{{\rm d}D_{\rm d}}
\def\dDs{{\rm d}D_{\rm s}}
\def\Msol{\rm M_\odot}
\def\kms{{\rm \ kms}}
\def\vrandl{v_{\rm rand,l}}
\def\vrandb{v_{\rm rand,b}}
\def\tp3{${t_{\rm E}}^3$}
\def\tm3{${t_{\rm E}}^{-3}$}
\def\rot{{\rm rot}}
\def\max{{\rm max}}
\def\kpc{{\rm kpc}}
\def\brk{{\rm brk}}
\def\Deff{D_{\rm eff}}
\def\dGamma{d\Gamma}
\def\p(v){{\rm p}(v)}
\def\dv{{\rm d}v}
\def\dM{{\rm d}M}
\def\dN{{\rm d}N}
\def\f{{\rm f}}

\newcommand\araa{{ARA\&A}} 
\newcommand\aj{{AJ}}       
\newcommand\aap{{A\&A}}    
\newcommand\apj{{ApJ}}     
\newcommand\apjl{{ApJL}}   
\newcommand\mnras{{MNRAS}} 
\newcommand\nat{{Nat.}}    

\title[]{Optical Depths and Time-Scale Distributions in Galactic Microlensing}
\author[Wood \& Mao]{Alexander Wood and Shude Mao
\thanks{{(awood, smao)@jb.man.ac.uk}} \\
Jodrell Bank Observatory, University of Manchester, Macclesfield, Cheshire SK11 9DL, UK
}

\date{Accepted ........ Received ........; in original form ........}
\pubyear{2005}

\begin{document}
\maketitle

\begin{abstract}
We present microlensing calculations for a Galactic model based on
\citet{HG03}, which is empirically normalised by star counts. We find
good agreement between this model and data recently published by the MACHO
and OGLE collaborations for the optical depth in various Galactic fields, 
and the trends thereof with Galactic longitude $l$ and latitude $b$. We 
produce maps of optical depth and, by adopting simple kinematic models, 
of average event time-scales for microlensing towards the Galactic bulge. 
We also find that our model predictions are in reasonable agreement with 
the OGLE data for the expected time-scale distribution. We show that the
fractions of events with very long and short time-scales due to a lens 
of mass $M$ are weighted by $M^2\, n(M)\, \dM$ and $M^{-1}\, n(M)\, \dM$
respectively, independent of the density and kinematics of the lenses.
\end{abstract}


\section{Introduction}
\label{sec:intro}

A microlensing event occurs when a luminous object is temporarily
magnified by a massive body, such as a star or dark matter object,
passing close to the line of sight and acting as a gravitational
lens. \citet{Pac86} advocated searching for microlensing events towards
the Large Magellanic Cloud in order to detect dark matter in the
Galactic halo. Soon three separate collaborations were conducting
systematic searches: OGLE \citep{Uda92}, EROS \citep{Aub93}, and MACHO
\citep{Alc93}, between them observing the Large and Small Magellanic
Clouds and the Galactic bulge. Other groups such as MOA 
(e.g. \citealt{Bon01}) have since joined the search, and thousands of 
microlensing events have now been detected (e.g., \citealt{Alc00a, 
Woz01, Sum03}), almost all towards the bulge. A much smaller number of 
microlensing candidates have also been identified toward the Large 
Magellanic Cloud (e.g., \citealt{Alc00b}) and M31 (e.g., \citealt{Nov05}). 
One of the main aims of all these observations is to accurately measure 
the optical depth, $\tau$ -- the probability of seeing a microlensing 
event at any given instant -- which can provide much information about 
the structure and mass distribution of the Galaxy and its halo.

Since the first estimates of $\tau$ by \citet{Pac91} and \citet{Gri91},
predictions based on increasingly refined models have consistently and
significantly disagreed with measurements based on increasingly large
sets of observational data. However, there are now signs of
convergence. \citet{HG03} -- hereafter HG03 -- used star counts from the
\emph{Hubble Space Telescope} (\emph{HST}) to normalise their Galactic
model, predicting $\tau = 1.63 \times 10^{-6}$ towards Baade's window
(BW), based on lensing of red clump giants (RCGs). They noted reasonable
agreement with two recent measurements towards the bulge, also based on
RCGs, of $\tau = 2.0\, (2.13) \pm 0.4 \times 10^{-6}$ and $\tau = 0.94\,
(1.08) \pm 0.30 \times 10^{-6}$, from the MACHO \citep{Pop01} and EROS
\citep{Afo03} collaborations, respectively. The numbers in parentheses
are from table 2 of \citet{Afo03}, who enabled a better comparison
between all bulge optical depth measurements to be made by adjusting the
values for their offset from BW. Now from 7 years of MACHO survey data,
\citet{Pop04} report $\tau = 2.17^{+0.47}_{-0.38} \times 10^{-6}$ at
$(l, b) = (1.50^\circ, -2.68^\circ)$, which is in excellent agreement
with recent theoretical predictions, including the \citeauthor{HG03}
result. Most recently, from the OGLE-II survey \citet{Sum05} find $\tau
= 2.37^{+0.53}_{-0.43} \times 10^{-6}$ at $(l, b) = (1.16^\circ,
-2.75^\circ)$, which is also consistent with the latest MACHO survey value.

In this paper we generate Monte Carlo simulations of the Galaxy based on
HG03. The outline of the paper is as follows.
\S\ref{sec:model} describes the model and theory, and
\S\ref{sec:results} presents our results: In \S\ref{sec:opdepth res} we
reproduce the HG03 $\tau_{\rm BW}$, and then compare our predicted
$\tau$ with the recent MACHO and OGLE results in various
directions. \S\ref{sec:maps} presents maps of optical depth and average
event time-scale (duration). These maps can be compared with observations
in any direction. In \S\ref{sec:tscales} we predict the event rate as a
function of time-scale and compare this to the distribution observed by
OGLE. In \S\ref{sec:frac} we show how at both long and short times the 
time-scale distribution is directly related to the lens mass function. 
We summarise our results in \S\ref{sec:discussion}.

\section{The Model}
\label{sec:model}

\subsection{Bulge and disc mass models}
\label{sec:models}

\citet{Dwe95} compared various hypothetical mass density models of the
bulge to the infrared light density profile seen by the Cosmic
Background Explorer (COBE) satellite. We use the G2 (barred) model from
their table 1, with $R_{\rm max}$ = 5 kpc. The bar is inclined by
$13.4^\circ$ to the Galactic centre line of sight, and the distance to
the Galactic centre is set at 8 kpc. \citeauthor{Dwe95} used 8.5 kpc, so
we adjust their model parameters accordingly. The model is then
normalised by \emph{HST} star counts (see the end of \S\ref{sec:populations}). 
This independent constraint can be used to normalise any bulge model.

For the disc, we use the local disc density model of \citet{Zhe01}, as 
extended to the whole disc by HG03. As the disc model is relatively secure 
(HG03), it will contribute only small uncertainties to predictions of the 
optical depth, so it is not renormalised as for the bulge model.

\subsection{Source and lens populations}
\label{sec:populations}

The optical depths reported by \citet{Pop04} are based on lensing of RCGs 
in the bulge, and HG03 assume only bulge RCG sources in their model. 
\citet{Sum05} observed lensing of red giants and red super giants as well 
as RCGs. We assume that these different types of stars follow the same bar 
density distribution and are bright enough to be seen throughout the bar, which 
corresponds to the case with $\gamma=0$ in the following eq. (\ref{eq:opdepth}).

Our lens mass function is generated as in HG03. Their unnormalised bulge 
mass function assumes initial star formation according to
\begin{equation}
\dN / \dM = k(M/M_\brk)^\alpha,
\label{eq:massfunc}
\end{equation}
where $M_\brk = 0.7\ \Msol$, $\alpha = -2.0$ for $M > M_\brk$, and 
$\alpha = -1.3$ for $M \leq M_\brk$, consistent with observations by 
\citet{Zoc00}. However HG03 extended this beyond the latter's lower limit of 
$M \sim 0.15\ \Msol$ to a brown dwarf cut-off of $M \sim 0.03\ \Msol$. We 
assume objects with masses 0.03--0.08 $\Msol$ and 0.08--1 $\Msol$ become brown 
dwarfs (BD) and main-sequence stars (MS) respectively, 1--8 $\Msol$ stars 
evolve into 0.6 $\Msol$ white dwarfs (WD), 8--40 $\Msol$ stars become 1.35 
$\Msol$ neutron stars (NS), and anything more massive forms a 5 $\Msol$ black 
hole (BH).

For MS stars we use the mass-luminosity relation of \citet{Cox99}, and take 
all other lenses to be dark. The model is then normalised by comparing 
extinction-adjusted MS counts to \emph{HST} star counts \citep{Hol98} as 
described in HG03. The same mass function and luminosity relation are also 
used for the disc. Strictly they should be independently estimated, but 
any uncertainties are small compared to others involved as we find disc 
stars account for only $\sim 20$ per cent of the total number of stars in BW.

\subsection{Kinematic model}
\label{sec:kinematics}

To calculate the event rate, we must also specify the velocities of the lenses, 
sources and observer. The observer velocity $v_{\rm O}$ is assumed to follow 
the Galactic rotation, so the two velocity components in $l$ and $b$ are given by
\begin{equation} 
v_{\rm O, l} = v_{\rm O,\rot} = 220 \kms^{-1}, \ \ \ \ \ \ v_{\rm O,b} = 0.
\label{eq:vo}
\end{equation}
The lens and source velocities in the $l$ and $b$ directions are given by
\begin{equation}
v_{\rm l} = v_\rot + \vrandl, \ \ \ \ \ \ v_{\rm b} = \vrandb,
\label{eq:v_lb}
\end{equation}
where the rotation velocity $v_\rot$ and the random velocity $v_{\rm rand}$ are 
from \citet{HG95}: for the disc $v_\rot = 220 \kms^{-1}$, and for the bar $v_\rot$ 
is given by projecting $v_\max = 100 \kms^{-1}$ across the line of sight according to
\begin{eqnarray}
v_\rot & = & v_\max \left( x \over 1\ \kpc \right) \ \ \ \ \ \ (R < 1\ \kpc,\ {\rm solid\ body\ rotation}),  \nonumber \\
v_\rot & = & v_\max \left( x \over R \right) \ \ \ \ \ \ \ \ \ \ \ (R \geq 1\ \kpc,\ {\rm flat\ rotation}),
\label{eq:vrot}
\end{eqnarray}
where $R=(x^2+y^2)^{1/2}$, and the coordinates $(x, y, z)$ have their origin at 
the Galactic centre, with the $x$ and $z$ axes pointing towards the Earth and 
the North Galactic Pole respectively. The random velocity components $\vrandl$ 
and $\vrandb$ are assumed to have Gaussian distributions. For the disc 
$\sigma_{\rm l,\, b} = (30,\,20) \kms^{-1}$, and for the bar we use 
$\sigma_{\rm x,\, y,\, z}$ = (110, 82.5, 66.3)$\kms^{-1}$ as found by \citet{HG95} 
using the tensor virial theorem (see also \citealt {SEW03}, and \citealt{KR02}). 
These values should be altered slightly as HG03 used a different normalisation. 
This may affect our results slightly, but it is re-assuring that our results 
based on such a simple kinematic model appear to agree with the data quite well 
(see \S\ref{sec:results}).

\subsection{Optical depth and event rate}
\label{sec:taugamma}

$\tau$ in any given direction is an average over the optical depths of all the 
source stars in that direction. The optical depth to a particular star is 
defined as the probability that it is within the Einstein radius (see below) 
of any foreground lenses. Hence more distant stars, although fainter and less 
likely to be detected, have higher optical depths \citep{Sta95}. HG03 accounted 
for this with the term $\gamma$ in the calculation of observed optical depth:
\begin{equation}
\langle \tau \rangle_\gamma = {4\pi G \over c^2} 
{\int_0^\infty \dDs \Ds^{2-\gamma} \rho(\Ds) \int_0^{\Ds} \dDd \rho(\Dd) \Dd (\Ds - \Dd)/\Ds \over \int_0^\infty \dDs \Ds^{2-\gamma} \rho(\Ds)},
\label{eq:opdepth}
\end{equation}
where $\Ds$ and $\Dd$ are the distances to the source and deflector (lens), 
and $\rho(\Ds)$ and $\rho(\Dd)$ are the source number density and lens mass 
density. RCGs and other bright stars in the bulge can be identified 
independently of their distance, so $\gamma = 0$. Eq. (\ref{eq:opdepth}) was 
originally presented (in a slightly different form) by \citet{KP94}, who also 
derived an expression for the lensing event rate $\Gamma$. We give this here 
in terms of $\gamma$, and account for variation in lens mass by bringing the 
term $M^{-1/2}$ inside the integral:
\begin{eqnarray}
\Gamma & = & {4 G^{1/2} \over c} \int_0^\infty \dDs \Ds^{2-\gamma} \rho(\Ds)  \nonumber \\
       &   & \times {\int_0^{\Ds} \dDd \rho(\Dd) v [\Dd (\Ds - \Dd)/M \Ds]^{1/2} \over \int_0^\infty \dDs \Ds^{2-\gamma} \rho(\Ds)},
\label{eq:freq}
\end{eqnarray}
where $v$ is the lens-source relative transverse velocity,
\begin{equation}
v = ({v_{\rm l}}^2 + {v_{\rm b}}^2)^{1/2},
\end{equation}
and its components in the Galactic $l$ and $b$ coordinates, $v_{\rm l}$ and 
$v_{\rm b}$, are related to the observer, lens and source velocities by
\begin{equation}
v_{\rm l, b} = \left( (v_{\rm D} - v_{\rm O}) + (v_{\rm O} - v_{\rm S}) {\Dd \over \Ds} \right)_{\rm l, b},
\end{equation}
where $v_{\rm D}$ and $v_{\rm S}$ are the deflector (lens) and source 
transverse velocities; their components in the $l$ and $b$ directions are 
given in eq. (\ref{eq:v_lb}).

The time-scale of an event $\tE$ is defined as the time taken for a source 
to cross the Einstein radius of the lens $r_E$ \citep{Pac96}:
\begin{equation}
\tE = {r_E \over v} \ \ \ \ \ \ \ \  r_E = \left( {4GM \over c^2} {\Dd(\Ds - \Dd) \over \Ds} \right)^{1/2}.
\label{eq:tE}
\end{equation}

\section{Results}
\label{sec:results}

\subsection{Optical depth in MACHO and OGLE fields}
\label{sec:opdepth res}

HG03 calculated $\tau = (0.98, 0.65, 1.63) \times 10^{-6}$ towards BW for bulge, 
disc, and all lenses respectively. Our equivalent values are (1.06, 0.65, 1.71) 
$\times 10^{-6}$. HG03 noted that the value of $\gamma$ makes little difference 
to $\tau$ for disc lenses, but for bulge lenses $\tau$ becomes $0.86 \times 10^{-6}$ 
when $\gamma = 1$. We find $\tau = 0.92 \times 10^{-6}$ in this case. Our results 
for bulge lenses differ by 7--8 per cent from HG03's due to a slight difference 
in implementation of the bulge model normalisation. We find that allowing MS disc 
lenses to also act as sources themselves makes a negligible difference to the 
total value of $\tau$.

The MACHO measurement \citep{Pop04} of $\tau = 2.17^{+0.47}_{-0.38} \times 10^{-6}$ 
at $(l, b) = (1.50^\circ, -2.68^\circ)$, was obtained from a sub-sample of their 
observed fields, the `Central Galactic Region' (CGR), which covers 4.5 deg$^2$ and 
contains 42 of the 62 RCG microlensing events seen. The coordinates 
$(1.50^\circ, -2.68^\circ)$ are a weighted average position of these fields; the 
unweighted average is $(l, b) = (1.55^\circ, -2.82^\circ)$. Optical depths were 
also given for a region `CGR+3' that contains 3 additional fields, and for all 62 
events. In Table \ref{tab:opdepths_macho} we compare our expected values to each 
of these results, and to $\tau$ reported for each of the individual CGR fields.

OGLE's measurement \citep{Sum05} of $\tau = 2.37^{+0.53}_{-0.43} \times 10^{-6}$ 
at $(l, b) = (1.16^\circ, -2.75^\circ)$ made use of 32 RCG events, in 20 of their 
49 fields, where $(l, b) = (1.16^\circ, -2.75^\circ)$ is the weighted average 
field position. $\tau$ was also given for each field; we compare our values to all 
of these results in Table \ref{tab:opdepths_ogle}.

Note that any significant disagreement occurs only in individual fields, and that 
in only 1 of the 6 fields (MACHO and OGLE) with $> 4$ events (OGLE \#30) does our 
value lie far outside the stated $1 \sigma$ uncertainty.

Table \ref{tab:frac_opdepth} shows the percentage contributions to the total optical 
depth and event rate from the different types of lenses. The disc lenses contribute 
about 37 per cent of the optical depth and a slightly smaller fraction (31 per cent) 
of the event rate. We see that 62 per cent of all events have luminous (MS) lenses, 
the other 38 per cent are dark (BD, WD, NS and BH). The NSs and BHs contribute about 
9 per cent of the optical depth but only 4 per cent of the event rate. This is because 
the events caused by stellar remnants on average have longer time-scales, and thus 
they occur less frequently. 

\begin{table}
\centering
\begin{tabular}{ccccc}\hline

  Region/field & ${N_{\rm events}}^*$ & ($l$, $b$) ($^\circ$) 
  & $\tau_{\rm MACHO} (\times 10^{-6})$ & $\tau_{\rm model} (\times 10^{-6})$ \\ \hline
 
  CGR$^\dag$   & 42  & (1.50, -2.68)  & $2.17^{+0.47}_{-0.38}$  & 2.43 \\
  CGR$^\ddag$  & 42  & (1.55, -2.82)  & --                      & 2.33 \\
  CGR+3        & 53  & (1.84, -2.73)  & $2.37^{+0.47}_{-0.39}$  & 2.34 \\
  All events   & 62  & (3.18, -4.30)  & $1.21^{+0.21}_{-0.21}$  & 1.32 \\
               &     &                &                         &      \\ 
  108          & 6   & (2.30, -2.65)  & $2.04 \pm 0.92$         & 2.31 \\
  109          & 2   & (2.45, -3.20)  & $0.58 \pm 0.41$         & 1.96 \\
  113          & 3   & (1.63, -2.78)  & $0.55 \pm 0.35$         & 2.34 \\
  114          & 3   & (1.81, -3.50)  & $1.19 \pm 0.74$         & 1.87 \\
  118          & 7   & (0.83, -3.07)  & $2.85 \pm 1.35$         & 2.25 \\
  119          & 0   & (1.07, -3.83)  & --                      & 1.74 \\
  401          & 7   & (2.02, -1.93)  & $5.13 \pm 2.16$         & 2.85 \\
  402          & 10  & (1.27, -2.09)  & $3.95 \pm 1.50$         & 2.89 \\
  403          & 4   & (0.55, -2.32)  & $1.16 \pm 0.66$         & 2.83 \\ \hline

\end{tabular}
\caption{Comparison of model and MACHO optical depths for the Central 
  Galactic Region (CGR) and individual fields. $^*$Number of events seen
  by MACHO. $^\dag$Weighted average ($l$, $b$). $^\ddag$Unweighted 
  average ($l$, $b$).}
\label{tab:opdepths_macho}
\end{table}

\begin{table}
\centering
\begin{tabular}{ccccc}\hline

  Region/field      & ${N_{\rm events}}^* $ & ($l$, $b$) ($^\circ$) 
  & $\tau_{\rm OGLE} (\times 10^{-6})$ & $\tau_{\rm model} (\times 10^{-6})$ \\ \hline

  All fields$^\dag$ & 32 & (1.16, -2.75)  & $2.37^{+0.53}_{-0.43}$ & 2.43 \\
                    &    &                &                        &      \\ 
  1                 & 0  & (1.08, -3.62)  & --                     & 1.87 \\
  2                 & 1  & (2.23, -3.46)  & $2.31 \pm 2.31$        & 1.85 \\
  3                 & 4  & (0.11, -1.93)  & $3.99 \pm 2.07$        & 3.20 \\
  4                 & 5  & (0.43, -2.01)  & $2.93 \pm 1.39$        & 3.09 \\
  20                & 1  & (1.68, -2.47)  & $1.15 \pm 1.15$        & 2.54 \\ 
  21                & 0  & (1.80, -2.66)  & --                     & 2.39 \\
  22                & 1  & (-0.26, -2.95) & $0.79 \pm 0.79$        & 2.42 \\
  23                & 0  & (-0.50, -3.36) & --                     & 2.13 \\
  30                & 6  & (1.94, -2.84)  & $8.88 \pm 3.89$        & 2.26 \\
  31                & 1  & (2.23, -2.94)  & $2.10 \pm 2.10$        & 2.15 \\
  32                & 1  & (2.34, -3.14)  & $0.87 \pm 0.87$        & 2.02 \\
  33                & 2  & (2.35, -3.66)  & $9.69 \pm 7.38$        & 1.73 \\
  34                & 2  & (1.35, -2.40)  & $3.80 \pm 2.69$        & 2.65 \\
  35                & 2  & (3.05, -3.00)  & $2.99 \pm 2.19$        & 1.98 \\
  36                & 0  & (3.16, -3.20)  & --                     & 1.85 \\
  37                & 2  & (0.00, -1.74)  & $2.06 \pm 1.65$        & 3.39 \\
  38                & 2  & (0.97, -3.42)  & $2.68 \pm 2.09$        & 2.01 \\
  39                & 3  & (0.53, -2.21)  & $1.51 \pm 0.90$        & 2.92 \\
  45                & 0  & (0.98, -3.94)  & --                     & 1.68 \\
  46                & 0  & (1.09, -4.14)  & --                     & 1.56 \\ \hline

\end{tabular}
\caption{Comparison of model and OGLE optical depths. $^*$Number of events 
  seen by OGLE. $^\dag$Weighted average ($l$, $b$).}
\label{tab:opdepths_ogle}
\end{table}

\begin{table}
\centering
\begin{tabular}{cccccccc}\hline

  & \multicolumn{7}{c}{Location/type of lens} \\ 
  & Bar & Disc & BD & MS & WD & NS & BH \\ \hline

  Optical depth & 63 & 37 & 7  & 62 & 22 & 6 & 3 \\
  Event rate    & 69 & 31 & 17 & 62 & 17 & 3 & 1 \\ \hline

\end{tabular}
\caption{Percentage contributions, to the total predicted $\tau$ and $\Gamma$, 
  from different types of lens.} 
\label{tab:frac_opdepth}
\end{table}

\begin{figure}
\centering
\includegraphics[width = 8cm]{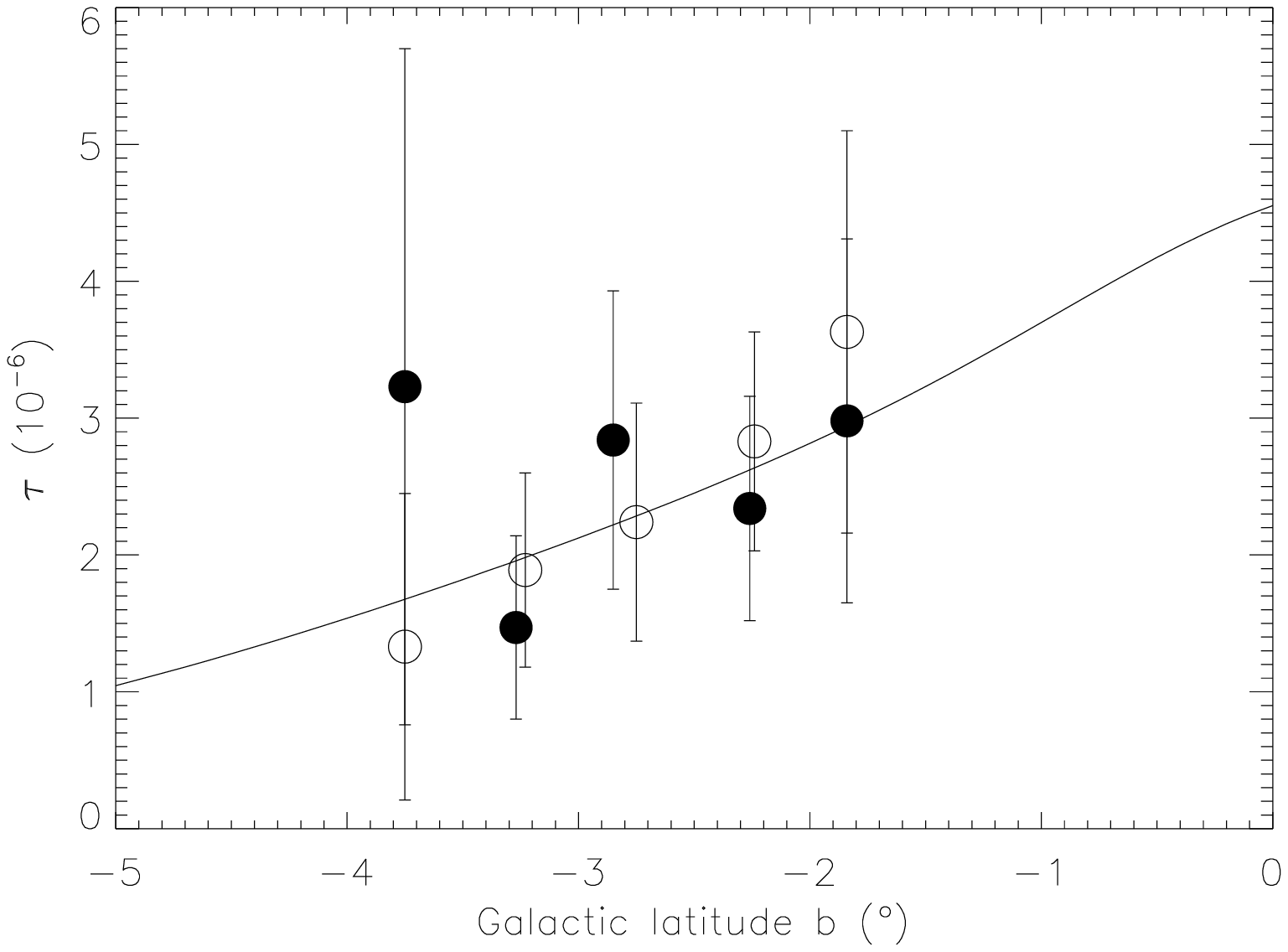}
\includegraphics[width = 8cm]{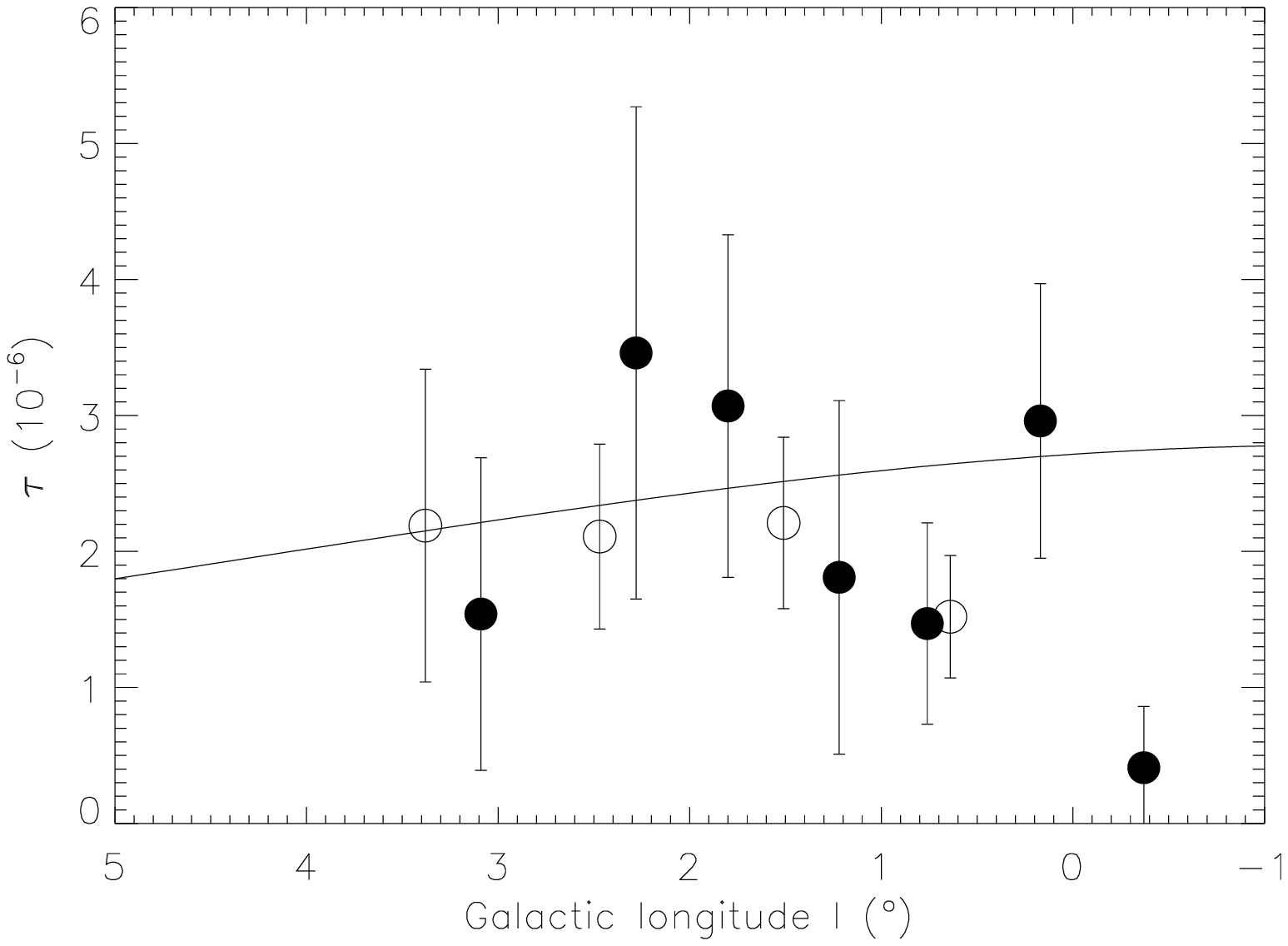}      
\caption{Average optical depth in latitude (left panel) and longitude (right 
  panel) strips, for $-5.5^\circ \leq b \leq 5.5^\circ$ and 
  $-5.5^\circ \leq l \leq 5.5^\circ$ respectively. The solid line shows the 
  model prediction, while the open and solid circles are data points from MACHO 
  (fig. 14, \citealt{Pop04}) and OGLE (fig. 12, \citealt{Sum05}) respectively.}
\label{fig:line}  
\end{figure}

In their figs. 12 and 14 respectively, \citet{Sum05} and \citet{Pop04} plot 
average optical depths in latitude and longitude strips. We produce similar 
plots in Fig. \ref{fig:line}, with the OGLE and MACHO data points shown. In 
both sets of strips the model is in good agreement with both sets of data. 
The single data point at negative $l$ is based on only one microlensing event, 
so the discrepancy has low statistical significance.

\subsection{Maps of optical depth and average event time-scale}
\label{sec:maps}

\begin{figure}
\centering
\includegraphics[width = 10cm]{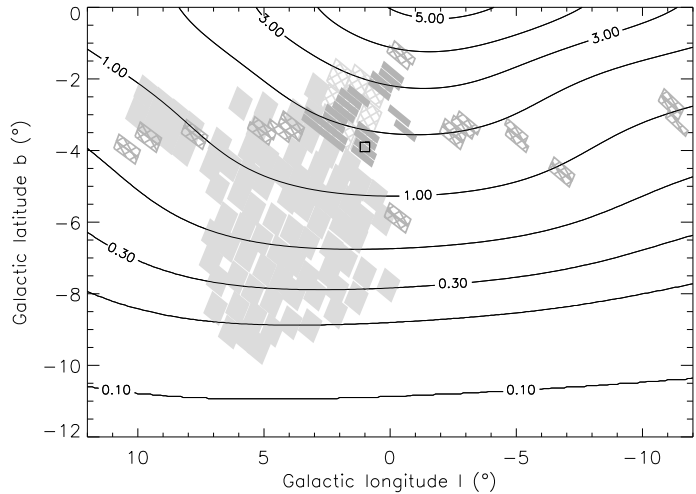}              
\caption{Map of expected optical depth. The MACHO and OGLE fields are shown 
  by the large and small grey boxes, respectively. For the MACHO fields, 
  the crosshatch pattern indicates the CGR subset listed in Table 
  \ref{tab:opdepths_macho}. For OGLE, the crosshatch pattern denotes those 
  fields \emph{not} listed in Table \ref{tab:opdepths_ogle}. The small 
  square indicates BW. Contour levels are at $(0.1, 0.2, 0.3, 0.5, 1, 2, 
  3, 4, 5) \times 10^{-6}$.}
\label{fig:map_opdepth}  
\end{figure}

\begin{figure}
\centering 
\includegraphics[width = 10cm]{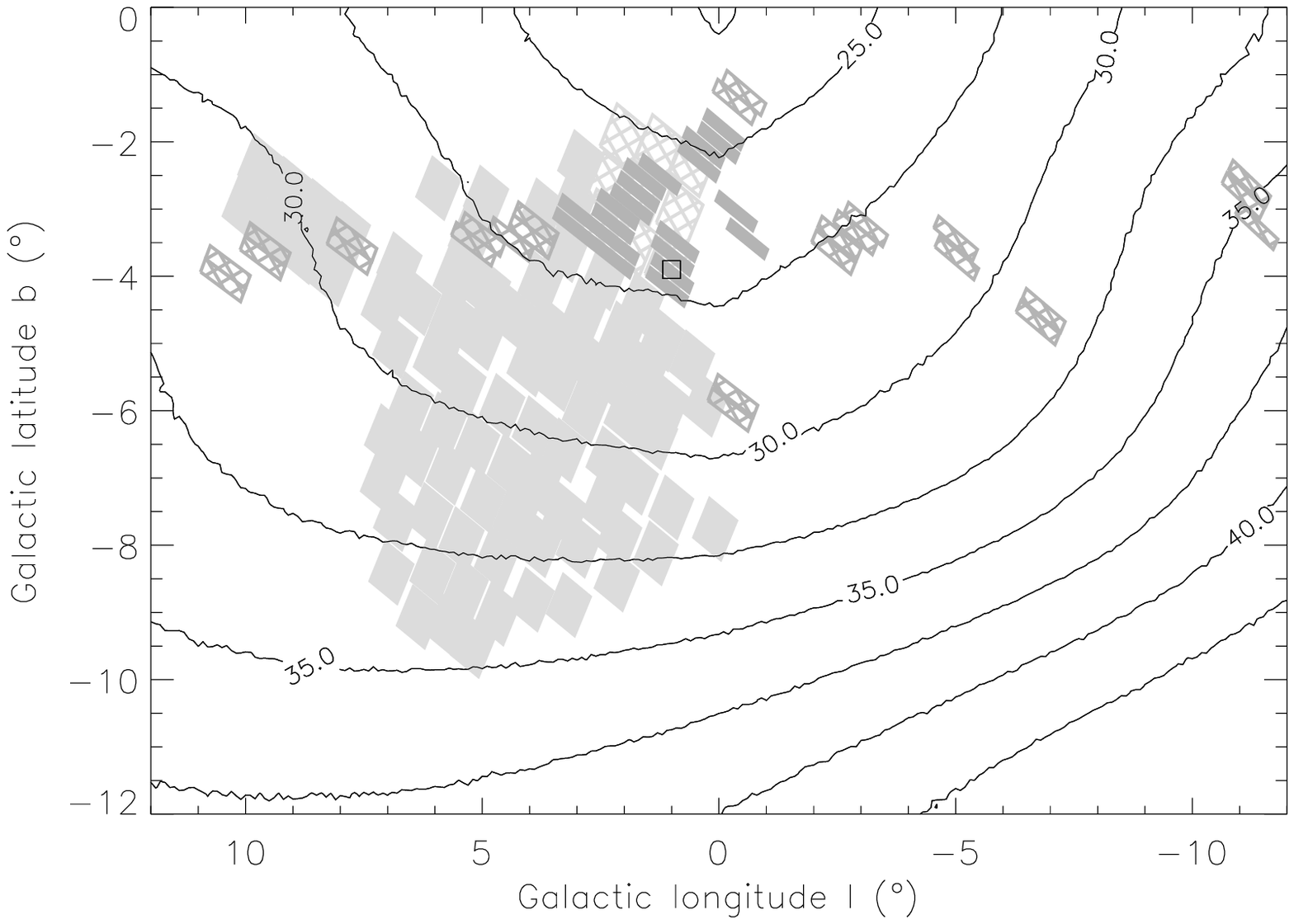}
\caption{Map of expected average event time-scale. The MACHO and OGLE fields 
  are shown by the large and small grey boxes, respectively. For the MACHO 
  fields, the crosshatch pattern indicates the CGR subset listed in Table 
  \ref{tab:opdepths_macho}. For OGLE, the crosshatch pattern denotes those 
  fields \emph{not} listed in Table \ref{tab:opdepths_ogle}. The small square 
  indicates BW. Contour levels are at 22.5, 25, 27.5, 30, 32.5, 35, 37.5, 40 
  and 42.5 d.}
\label{fig:map_time}  
\end{figure}

Figs. \ref{fig:map_opdepth} and \ref{fig:map_time} are maps of expected optical 
depth and average event time-scale. We can clearly see higher optical depths 
and longer time-scales at negative galactic longitude. This is due to the 
inclination of the bar to the line of sight. At positive longitude the bar is 
closer to us, and the line of sight cuts through the bar at a steeper angle. 
Hence there are fewer potential lenses, in either the disc or the bar, between 
us and any bar source, and so $\tau$ is smaller. Also, objects rotating around 
the Galactic centre have a smaller component of their velocity along the line 
of sight, so average transverse velocities will be greater, and average 
time-scales shorter. At negative longitude, the line of sight passes through 
more of the disc and cuts the bar at a shallower angle. Hence we see higher 
optical depths and smaller transverse velocities, and thus longer average 
time-scales.

\begin{figure}
\centering 
\includegraphics[width = 10cm]{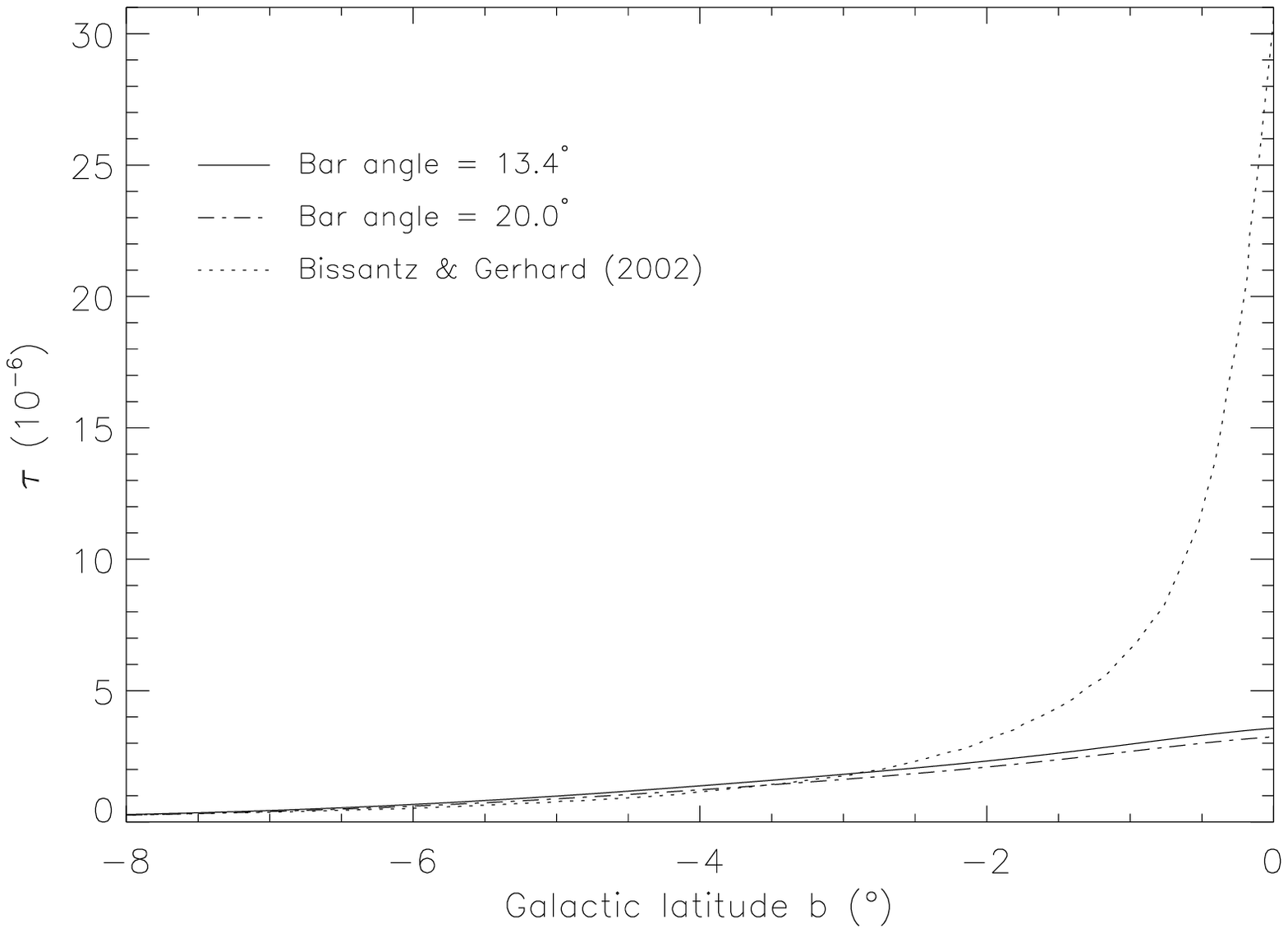}
\caption{$\tau$ as a function of $b$, for $l$ = $3.9^\circ$. Our model slope is 
  almost identical for bar angles of $13.4^\circ$ and $20^\circ$. The profile of 
  \citet{BG02} diverges from ours at $b\approx -3^\circ$, increasing rapidly 
  towards the mid-plane ($b=0^\circ$).}
\label{fig:gerhard}  
\end{figure}

We wish to compare our maps to others. \citet{EB02} produced red clump optical 
depth maps for three Galactic models, but while two of these appear similar to 
ours, they do not agree. One of those models was also used to make a time-scale 
map, which is quite different to ours. This is not surprising since, 
as well as using a different mass model, their mass function, velocities and 
velocity dispersions were also different. (In fact their timescale map has two 
sets of contours, to show the effects of including and excluding bar streaming. 
Without streaming their mean timescales are much shorter than ours, and with it 
they are greater by a factor $\sim3$, much longer than ours. Such a large 
variation is puzzling, and we are cautious about comparing their map to ours).

In their fig. 16, \citet{BG02} presented an optical depth map for RCG sources, 
with a bar angle of $20^\circ$. For $b$ $\la -3^\circ$ it appears quite similar 
to ours, but moving towards the Galactic centre $\tau$ climbs far more steeply 
than in our map. This is best seen by comparison to their fig. 17, where they 
plot $\tau$ as a function of $b$, for $l$ = $3.9^\circ$. This is shown in 
Fig. \ref{fig:gerhard} with equivalent profiles for our model. We see how 
rapidly \citeauthor{BG02}'s profile diverges from ours towards $b=0^\circ$. We 
also see that changing the bar angle in our model from $13.4^\circ$ to $20^\circ$ 
does not explain this difference. Instead it is probably due to the density in 
their bulge mass model increasing much faster towards the mid-plane. The 
observational data for the mid-plane are limited due to heavy extinction, and so 
mass models are not well-constrained in this region. Given the difficulty in 
obtaining any measurement of $\tau$ at small latitude, it is difficult at present 
to test either profile there.

\subsection{Time-scale distributions}
\label{sec:tscales}

\begin{figure}
\centering
\includegraphics[width = 10cm]{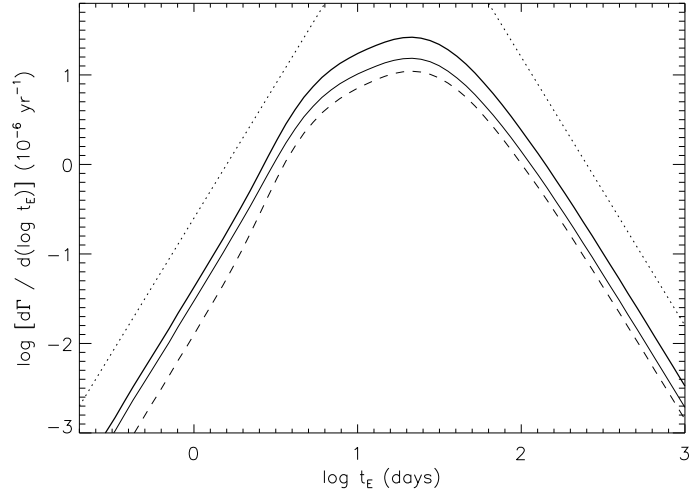}
\caption{Predicted microlensing event rate as a function of time-scale, towards 
  $(l, b) = (1.16^\circ, -2.75^\circ)$. The bold line represents all lenses. The 
  thin and dashed lines represent the bar and disc lenses, respectively. The 
  two dotted lines are asymptotic tails $\dGamma / d({\rm log}\ \tE) \propto$ 
  \tp3, \tm3 for very short and long time-scales, respectively.}
\label{fig:tscales}
\end{figure}

Fig. \ref{fig:tscales} shows the event rate as a function of time-scale towards 
the OGLE coordinates $(l, b) = (1.16^\circ, -2.75^\circ)$, for bar (thin line), 
disc (dashed line) and all (bold line) lenses. There is good agreement with the 
asymptotic power-law tails $\dGamma / d({\rm log}\ \tE) \propto$ \tp3, \tm3 for 
very short and long time-scales, respectively \citep{MP96}. The disc lensing 
events have an average time-scale of 26.3 d, slightly longer than the bulge 
lensing events' average of 25.7 d, as also found by \citet{KP94}. The average 
time-scale for all events is 25.9 d.

\begin{figure}
\centering
\includegraphics[width = 10cm]{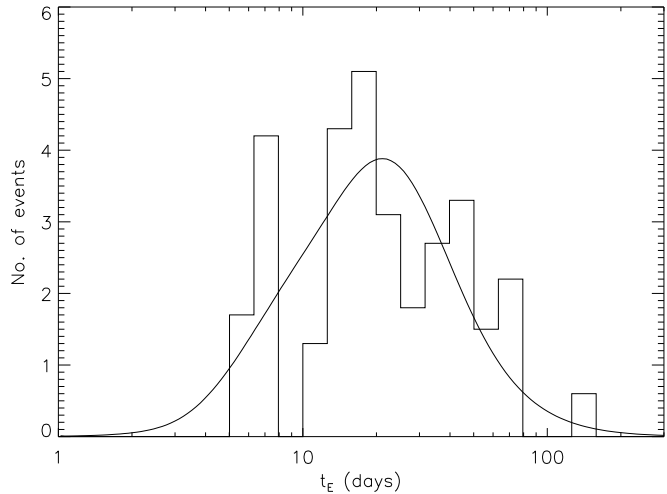}
\caption{Microlensing event rate as a function of time-scale, towards 
  $(l, b) = (1.16^\circ, -2.75^\circ)$. The solid line shows the model 
  prediction, and the OGLE observed distribution (corrected for detection 
  efficiency) is shown as a histogram.}
\label{fig:tscales2}
\end{figure}

In Fig. \ref{fig:tscales2} we renormalise our time-scale distribution (for 
all lenses) and compare it to that seen by OGLE, as corrected for detection 
efficiency (see fig. 14 in \citealt{Sum05}). We do not compare to the 
time-scale distribution seen by \citet{Pop04} -- they assumed that the effect 
of blending on RCG sources is negligible, but \citet{Sum05} found $\approx$
38 per cent of OGLE-II events with apparent RCG sources were really due to
faint stars blended with a bright companion. Fortunately, they also showed 
that blending has little effect on estimates of $\tau$ due to partial 
cancellation of its different effects, a point also made by \citet{Pop04}. 
However, time-scale distributions will be significantly shifted towards 
shorter events. As a result, the MACHO time-scale distribution (not shown) 
has a significant excess at short time-scales compared with our model.

Our time-scale distribution shows reasonable agreement with OGLE's. The 
Kolmogorov--Smirnov (KS) test shows that the predicted and observed distributions 
are consistent at a $\approx 52$ per cent confidence level. Our average time-scale 
of 25.9 d is in excellent agreement with OGLE's corrected average of 
$28.1 \pm 4.3$ d. Our median and quartiles are (19.2, 11.2, 31.7) d, respectively. 

The event time-scale distribution from the data still has large uncertainties due 
to the limited number of events. It is apparent that the data have not yet 
reached the predicted asympototic behaviour at short and long time scales, so a 
more stringent test on the model is not yet possible.

\citet*[see also \citealt{Pea98}]{BDG04} have also modelled the time-scale 
distribution. They reproduced that from MACHO's 99 DIA events \citep{Alc00a} 
centred at $(l, b) = (2.68^\circ, -3.35^\circ)$. However, both distributions are 
clearly shifted towards short time-scales compared to our model prediction in the 
same direction\footnote{At first glance all three distributions may appear to be 
similar. However, whereas we define the event timescale as the Einstein-radius 
crossing time (see \S\ref{sec:taugamma}), MACHO plot the diameter-crossing time, 
a factor of 2 difference.} (this is not shown, as it is very close to the solid 
line in Fig. \ref{fig:tscales2}). Although the DIA method is less prone to the 
systematics of blending \citep{Sum05}, it is still possible that the MACHO DIA 
time-scale distribution is somewhat affected. The most important difference 
between our model and \citeauthor{BDG04}'s is that in order to match the data at 
short time-scales, they adopted a Schecter mass function, $n(M) \propto M^{-2.35}$ 
for $M \leq 0.35\ \Msol$ down to 0.04 $\Msol$, steeper than our mass function, 
$n(M) \propto M^{-1.3}$ for $M < 0.7\ \Msol$. As a result, their median lens mass 
is much smaller than ours (0.11 $\Msol$ vs. 0.35 $\Msol$, weighted by event rate). 
The different kinematics may also have a noticeable effect on the timescales, but 
their more realistic dynamical model does not allow a simple comparison to be made.

\subsection{Fractional contributions to event rate -- mass weightings}
\label{sec:frac}

\begin{figure}
\centering
\includegraphics[width = 10cm]{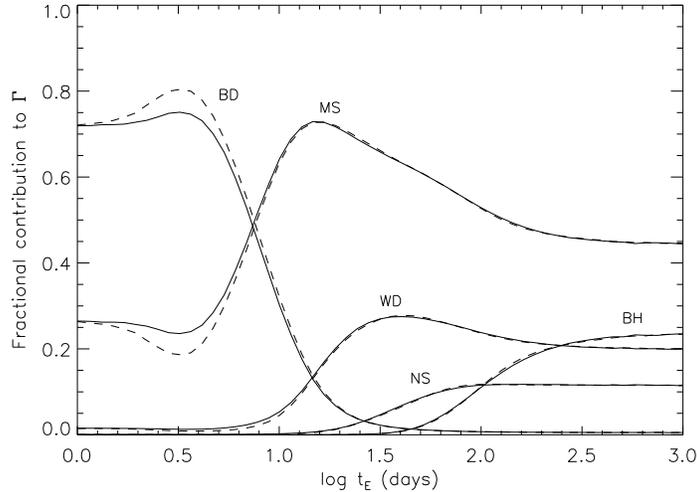}                          
\caption{Fractional contributions to total expected event rate, as a function of event 
  time-scale, from BD, MS, WD, NS and BH lenses as indicated. The solid and dashed lines 
  represent the bar and disc lenses, respectively. The asymptotic fractions at long and 
  short time-scales are a function of the lens mass only (see text).}
\label{fig:frac_tscales}
\end{figure}

Fig. \ref{fig:frac_tscales} shows the fractional contributions to the total event 
rate, as a function of event time-scale, for the different types of lens (BD, MS, 
WD, NS and BH) as indicated. At short time-scales ($\tE \la 4$ d), the brown dwarfs 
dominate the event rate, while at long time-scales ($\tE \ga 100$ d), the stellar 
remnants become increasingly important. There is asymptotic behaviour at both long 
and short time-scales. We find that the fractional contribution from a lens of mass 
$M$ is weighted by $M^2\, n(M)\, \dM$ and $M^{-1}\, n(M)\, \dM$, respectively. In 
the Appendix we derive these weightings from eq. (\ref{eq:freq}). (The scaling at 
long event tails has already been derived by \citealt{Ago02}). Table 
\ref{tab:frac_tscales} shows that direct calculation of these asymptotic fractions 
from the mass function gives results that clearly agree with the trends in Fig. 
\ref{fig:frac_tscales}. These weightings are independent of the density and kinematics 
of the lens population, and hence provide valuable information about the lens mass 
function.

\begin{table}
\centering
\begin{tabular}{cccccc}\hline

  Time-scale & BD   & MS & WD  & NS    & BH     \\ \hline
  Long       & 0.53 & 44 & 20  & 12    & 24     \\
  Short      & 72   & 27 & 1.5 & 0.078 & 0.0032 \\ \hline

\end{tabular}
\caption{Percentage contributions to the total predicted event rate, at long and 
  short time-scales, from the different types of lens.} 
\label{tab:frac_tscales}
\end{table}

\section{Summary}
\label{sec:discussion}

In this paper, we have used a simple Galaxy model normalised by star counts \citep{HG03} 
to predict the microlensing optical depth. Combined with simple kinematic models, we 
also predict maps and distributions of the time-scale distributions. We have shown that 
the fraction of long and short events contributed by a lens of mass $M$ is weighted by
$M^2\, n(M)\, \dM$ and $M^{-1}\, n(M)\, \dM$ respectively. If the tails of this 
distribution can be accurately determined from observations, we have a direct probe of 
the lens mass function.

It is remarkable that this emprically-normalised model based on the COBE G2 model 
\citep{Dwe95} shows good agreement with data recently published by the MACHO and OGLE 
collaborations (\citealt{Sum05} and \citealt{Pop04}) for the optical depth in various
Galactic fields, and its trends with $l$ and $b$. Our maps of optical depth and average 
event time-scale cover a large area of the sky, and can be compared to future 
determinations of $\tau$ in similar areas when they become available. The expected 
distribution of the event time-scale also appears to show good agreement with the 
recently published OGLE data \citep{Sum05}. However, the numbers of microlensing events 
used (42 and 32) in the recent MACHO and OGLE analyses are still small, so the test 
on the models is not yet stringent. When the much larger database of microlensing 
events ($\sim$ thousands) is analysed, then a full comparison with the models will 
become much more discriminating.

\section*{Acknowledgments}

We thank Drs. Vasily Belokurov, Nicholas Rattenbury and Martin Smith for many useful 
discussions. We thank the anonymous referee for their helpful comments. AW acknowledges 
support from a PPARC studentship.

\bibliographystyle{mn2e}

\appendix
\label{app:assym}

\section{Event rate weightings at long and short time-scales}

As described in \S\ref{sec:frac}, the microlensing event rate shows asymptotic behaviour 
at both long and short time-scales. We show here that this is directly related to the lens 
mass function, specifically, the fractional contributions are weighted by $M^2\, n(M)\, \dM$ 
and $M^{-1}\, n(M)\, \dM$, at very long and short time-scales respectively.

\label{app:assym intro}

The event rate is given by eq. (\ref{eq:freq}). However, as the mass dependence of the 
asympototic behaviour is the same for sources at different distances, we shall ignore 
the source distance dependences here. Therefore for a source at distance $\Ds$ and a 
population of lenses each with identical velocity $v$ and mass $M$, the event rate is 
given by
\begin{equation}
\Gamma = {4G^{1/2} \over c} \int_0^{\Ds} \rho(\Dd)\, v \left({\Dd (\Ds - \Dd) \over M \Ds} \right) ^{1/2} \dDd,
\label{dfreq1}
\end{equation}
where $\rho(\Dd)$ is the lens mass density at $\Dd$.

In reality, $v$ and $M$ both vary. The velocity probability distribution, $\p(v)\, \dv$, 
can usually be approximated by a two-dimensional Maxwellian distribution
\begin{equation}
\p(v)\, \dv = {1 \over \sigma^2} \exp(-v^2 / 2 \sigma^2)\, v\, \dv,
\end{equation}
where $\sigma$ is the velocity dispersion. For constant $M$, the factor $\rho(\Dd)$ in 
eq. (\ref{dfreq1}) is simply the total mass density. When $M$ varies, the event rate depends 
on the lens mass function, i.e. on how the total mass is partitioned into lenses of different 
masses. We assume that this is the same everywhere. The mass density for lenses with 
$M \rightarrow M +\, \dM$ can be written as a product of $\f(\Dd)\, n(M)\, \dM$, where $\f(\Dd)$ 
indicates the distance dependence and $n(M)\dM$ is the number density of lenses between 
$M \rightarrow M +\, \dM$.

Integrating over the mass and velocity distributions and using the fact that 
$\rho(M)\dM = M \f(\Dd)\, n(M)\, \dM$, we obtain
\begin{equation}
\Gamma = 2 A^{1/2} \int_0^{\Ds} {\Deff}^{1/2} \f(\Dd)\, \dDd \int n(M) M^{1/2}\, \dM \int v\, \p(v)\, \dv,
\label{dfreq2}
\end{equation}
where $A = {4G / c^2} {\rm \ and\ } \Deff = \Dd (\Ds - \Dd) / \Ds$. We can now rewrite the 
time-scale equation (eq. \ref{eq:tE}) 
\begin{eqnarray} 
\tE = {r_E \over v} & = & \left( {4GM \over c^2} {\Dd (\Ds - \Dd) \over \Ds} \right) ^{1/2} v^{-1}  \nonumber \\
\label{tE2}
                    & = & {A^{1/2} M^{1/2} {\Deff}^{1/2} \over v},
\label{tE3}
\end{eqnarray} 
The typical transverse velocity is $\sim \sigma$, and this defines a characteristic time-scale as
\begin{equation}
t_\sigma = {A^{1/2} M^{1/2} {\Deff}^{1/2} \over \sigma}.
\end{equation}
The short and long tails satisfy $\tE \ll t_\sigma$ and $\tE \gg
t_\sigma$, respectively.

\subsection{Behaviour at long time-scales}
\label{app:assym long}

As can be seen from eq. (\ref{tE3}), the long time-scale events occur when the lens and source 
both move approximately parallel to each other and perpendicular to the line of sight. In this 
case, the transverse velocity is close to zero ($v \ll \sigma$) and the time-scale becomes long.

For events with time-scales longer than $t_{\rm long} (\gg t_\sigma)$, the transverse 
velocity must satisfy
\begin{equation}
v < {A^{1/2} M^{1/2} {\Deff}^{1/2} \over t_{\rm long}} \ll \sigma.
\label{smallv}
\end{equation}
The exponential factor $\exp(-v^2 / 2 \sigma^2)$ approaches unity, and so we have
\begin{eqnarray}
\Gamma (> t_{\rm long}) & = & {2 A^{1/2} \over \sigma^2} \int_0^{\Ds} {\Deff}^{1/2} \f(\Dd)\,                                                                                               \dDd \int n(M) M^{1/2}\, \dM \int_0^{A^{1/2} M^{1/2} {\Deff}^{1/2}                                                                                            \over t_{\rm long}} v^2 \dv  \nonumber \\
                        & = & {2 A^{1/2} \over \sigma^2} \int_0^{\Ds} {\Deff}^{1/2} \f(\Dd)\,                                                                                               \dDd \int n(M) M^{1/2}\, \dM \times {1 \over 3} \left({A^{3/2}                                                                                                M^{3/2} {\Deff}^{3/2} \over {t_{\rm long}}^3} \right)  \nonumber \\
                        & = & {2 A^2 \over 3 \sigma^2 {t_{\rm long}}^3} \int_0^{\Ds} {\Deff}^2                                                                                              \f(\Dd)\, \dDd \int n(M) M^2\, \dM.
\label{dfreq4}
\end{eqnarray}
Therefore, for long time-scale events, the event rate follows a power-law as a function of 
time-scale, with a normalisation that depends on the mass function, $\propto M^2\, n(M)\, \dM$, 
as also derived by \citet{Ago02}.

\subsection{Behaviour at short time-scales}
\label{app:assym short}

Re-expressing equation (\ref{tE2}) in terms of $x = \Dd / \Ds$, we have
\begin{equation} 
\tE = \left( {4GM \over c^2} x (1 - x) \Ds \right) ^{1/2} v^{-1}.
\label{tE5}
\end{equation} 
Very short events occur when the lens is very close to either the source or the observer, i.e., 
when $x \rightarrow 1$ or $x \rightarrow 0$. The asymptotic behaviour is the same for $x \ll 1$ 
and $1 - x \ll 1$, so we concentrate here on the case when $x \ll 1$, $x (1 - x) \approx x$. So 
for events shorter than a given time-scale $t_{\rm short} (\ll t_\sigma)$, we must have
\begin{equation} 
x < {v^2 {t_{\rm short}}^2 \over A\, M\, \Ds}.
\label{smallx}
\end{equation} 
Equation (\ref{dfreq2}) can then be re-written in terms of $x$:
\begin{equation}
\Gamma (<t_{\rm short}) = 2 A^{1/2} \int_0^{\Ds} [x\, (1 - x) \Ds]^{1/2}\, \f(\Dd)\, \dDd                                                                                               \int n(M) M^{1/2}\, \dM \int v\, \p(v)\, \dv.
\label{dfreq5}
\end{equation}
Changing the integration variable to $x$, and with
$x \ll 1$,  $\f(x\Ds)\approx \f(0)$, we obtain for the first integral
\begin{eqnarray}
\Gamma(<t_{\rm short}) & = & 2 A^{1/2} \int_0^{{v^2 t_{\rm short}}^2 \over A\, M\, \Ds} x^{1/2}\,                                                                                          \f(x \Ds) {\Ds}^{3/2}\, dx \int n(M) M^{1/2}\, \dM \int  v\, \p(v)\,                                                                                          \dv  \nonumber \\
                       & = & {4 \over 3} {{t_{\rm short}}^3 \over A} \f(0) \int M^{-1} n(M)\,\,                                                                                            \dM \int v^4 \, \p(v)\, \dv.
\label{dfreq6}
\end{eqnarray}
Therefore for short time-scale events, the event rate follows a power-law as a \nobreak{function} of 
the time-scale, with a normalisation that depends on the mass function, $\propto M^{-1} n(M)\, \dM$.

\end{document}